\documentclass[aps,twocolumn,showpacs,superscriptaddress,amsmath,amssymb,amsfonts,floatfix,longbibliography]{revtex4-2}

\usepackage[T1]{fontenc} 
\usepackage{graphicx}
\usepackage{float}
\usepackage{dcolumn}
\usepackage{bm}
\usepackage{amssymb}
\usepackage{tabularx}
\usepackage{microtype}
\usepackage{xfrac}
\usepackage{chemformula}
\usepackage{amsmath}
\usepackage{gensymb}
\usepackage[most]{tcolorbox}
\usepackage{xcolor}
\usepackage{multirow}
\usepackage{enumitem}
\usepackage{array}
\usepackage{booktabs}
\usepackage[flushleft]{threeparttable}
\usepackage{natbib,hyperref}
\usepackage{graphicx}
\usepackage{makecell}
\usepackage{colortbl}
\usepackage{xcolor}
\usepackage{soul}
\usepackage{braket}

\newcommand{\angstrom}{\mbox{\normalfont\AA}}

\hypersetup{
	colorlinks=true, 
	citecolor=blue,  
}

\begin{document}

\title{\textcolor{black}{Vanishing ordered moment in the frustrated triangular lattice antiferromagnet CuNdO$_2$}}

\author{Jonathan Gaudet}
    \affiliation{NIST Center for Neutron Research, National Institute of Standards and Technology, Gaithersburg, MD 20899, USA}
    \affiliation{Department of Materials Science and Eng., University of Maryland, College Park, MD 20742, USA}

\author{Dalmau Reig-i-Plessis}
    \affiliation{Stewart Blusson Quantum Matter Institute, University of British Columbia, Vancouver, BC V6T 1Z4, Canada}
    \affiliation{Laboratory for Solid State Physics, ETH Zurich, 8093, Zurich, Switzerland}

\author{Bogeng Wen}
    \affiliation{Department of Physics, University of Toronto, Toronto, ON M5S 1A7, Canada}

\author{Thomas J. Hicken}
    \affiliation{PSI Center for Neutron and Muon Sciences, 5232 Villigen PSI, Switzerland}

\author{Jonas A. Krieger}
    \affiliation{PSI Center for Neutron and Muon Sciences, 5232 Villigen PSI, Switzerland}

\author{Jan Peter Embs}
    \affiliation{PSI Center for Neutron and Muon Sciences, 5232 Villigen PSI, Switzerland}

\author{Hubertus Luetkens}
    \affiliation{PSI Center for Neutron and Muon Sciences, 5232 Villigen PSI, Switzerland}

\author{Adam A. Aczel}
    \affiliation{Neutron Scattering Division, Oak Ridge National Laboratory, Oak Ridge, TN 37831, USA}

\author{Stuart A. Calder}
    \affiliation{Neutron Scattering Division, Oak Ridge National Laboratory, Oak Ridge, TN 37831, USA}

\author{Matthew B. Stone}
    \affiliation{Neutron Scattering Division, Oak Ridge National Laboratory, Oak Ridge, TN 37831, USA}

\author{Hae-Young Kee}
\thanks{Email: \href{mailto:hy.kee@utoronto.ca}{hy.kee@utoronto.ca} and \href{mailto:alannah.hallas@ubc.ca}{alannah.hallas@ubc.ca}}
    \affiliation{Department of Physics, University of Toronto, Toronto, ON M5S 1A7, Canada}
    \affiliation{Canadian Institute for Advanced Research (CIFAR), Toronto, ON M5G 1M1, Canada}

\author{Alannah M. Hallas}
\thanks{Email: \href{mailto:hy.kee@utoronto.ca}{hy.kee@utoronto.ca} and \href{mailto:alannah.hallas@ubc.ca}{alannah.hallas@ubc.ca}}
\affiliation{Stewart Blusson Quantum Matter Institute, University of British Columbia, Vancouver, BC V6T 1Z4, Canada}
    \affiliation{Department of Physics \& Astronomy, University of British Columbia, Vancouver, BC V6T 1Z1, Canada}
    \affiliation{Canadian Institute for Advanced Research (CIFAR), Toronto, ON M5G 1M1, Canada}

\date{\today}
\begin{abstract}

We investigate the magnetic ground state of CuNdO$_2$, which is a delafossite with a triangular lattice of magnetic Nd$^{3+}$ ions that are well separated by non-magnetic Cu spacer layers. From inelastic neutron scattering measurements of the crystal electric field, we determine the strong Ising character of the pseudo-spin $\sfrac{1}{2}$ Nd$^{3+}$ moments. Magnetic susceptibility and heat capacity measurements reveal the onset of long-range antiferromagnetic order at $T_N=0.78$~K. While the magnetic transition is definitively observed with muon spin relaxation, accompanied by the formation of a weakly dispersing spin wave excitation, no dipole-ordered moment is detected with neutron diffraction. We show that the apparent absence of a dipolar ordered moment is a consequence of the dominant Ising character of the \textcolor{black}{antiferromagnetically coupled} Nd$^{3+}$ moments, which experience extreme frustration on the triangular lattice. Consequently, the frustration in CuNdO$_2$ is relieved through in-plane ordering of the substantially smaller perpendicular component of the Nd$^{3+}$ moments into a 120\textdegree\ structure, with a nearly vanishing ordered moment.

\end{abstract}

\maketitle

\section*{\label{sec:Introduction}Introduction}

Triangular lattices are one of the most ubiquitous lattice geometries in solid-state materials. 
When occupied by a magnetic cation, these triangular lattices are prone to experience competing magnetic interactions, which can complicate or even suppress the onset of long-range magnetic order, a phenomenon known as geometric frustration~\cite{greedan2001geometrically,balents2010spin,broholm2020quantum}.
In order to resolve its frustration, the compromise ground state of antiferromagnetically coupled Heisenberg spins on a triangular lattice is the so-called 120\textdegree\  structure~\cite{huse1988simple,jolicoeur1989spin,bernu1992signature,elstner1993finite,white2007neel}, an ordered state in which the spins in each triangular unit align with three-way mutual angles of 120\textdegree. However, the incorporation of additional spin couplings~\cite{misguich1999spin} and further neighbor exchange interactions~\cite{zhu2015spin,hu2015competing} can stabilize other ordered states and, in some cases, can even favor a quantum disordered spin liquid ground state. This Heisenberg limit is best realized in the case of insulating $3d$ transition metal compounds; however, in most cases, the frustration is partially alleviated by structural distortions to the triangular lattice~\cite{clarke1998synthesis,shimizu2003spin,ye2006spontaneous,mcqueen2008successive}.




Another set of outcomes can be observed for triangular lattice materials where the single-ion anisotropy deviates from the isotropic Heisenberg limit. This scenario occurs in the case of $4f$ rare-earth based materials, where substantial spin-orbit coupling, acting in conjunction with the crystal electric field, can generate both uniaxial (Ising) or planar (XY) spin anisotropies. Moreover, these anisotropic moments are themselves anisotropically coupled. Theoretical treatment of this type of triangular lattice material with anisotropic couplings gives rise to rich phase diagrams with a variety of ordered and spin liquid ground states~\cite{li2016anisotropic,li2016hidden,luo2017ground,iaconis2018spin,zhu2018topography,liu2018selective}. 
Many attempts to realize this phenomenology have centered on Yb as the magnetic cation, including NaYbO$_2$~\cite{bordelon2019field}, YbMgGaO$_4$~\cite{li2015rare,paddison2017continuous}, YbBO$_3$~\cite{somesh2023absence}, and YbZn$_2$GaO$_5$~\cite{bag2024evidence}, with other rare-earths yet to take center stage.

\begin{figure*}[htbp]
\centering
\includegraphics[width=\textwidth]{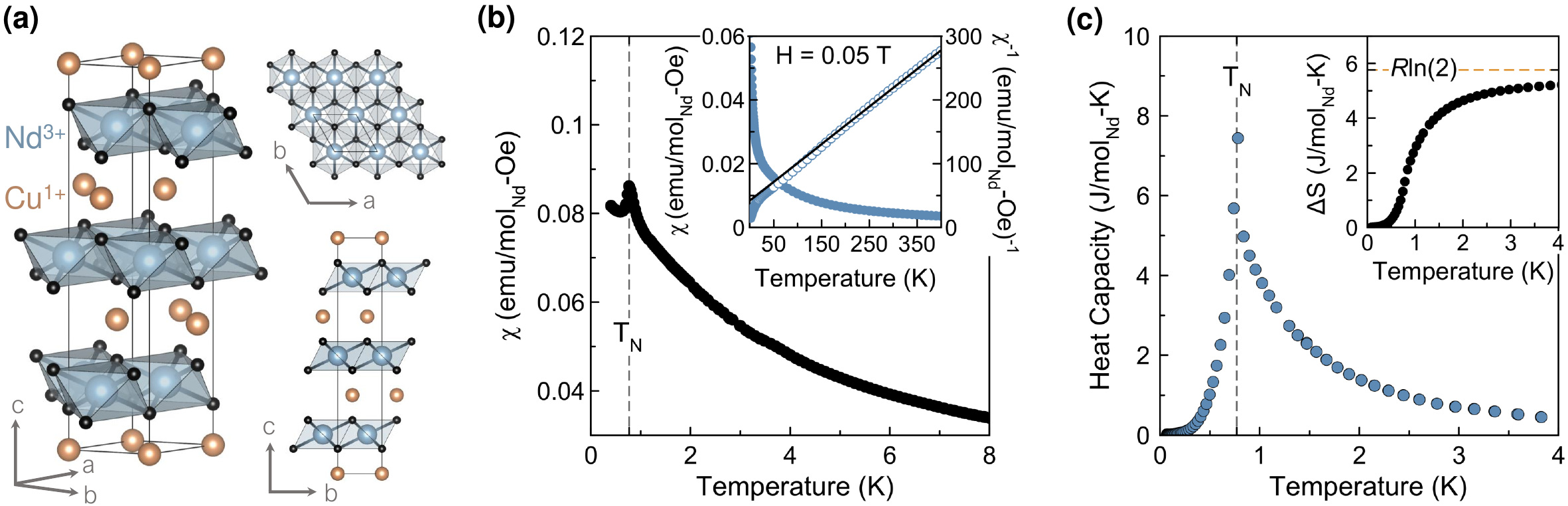}
 \caption{\textbf{Onset of long-range magnetic order in the triangular lattice antiferromagnet CuNdO$_2$} \textbf{(a)} The $R\overline{3}m$ crystal structure of CuNdO$_2$ is composed of alternating, staggered triangular layers of magnetic Nd$^{3+}$ separated by non-magnetic Cu$^{1+}$. \textbf{(b)} The magnetic susceptibility of CuNdO$_2$, measured in an $\mu_0H = 0.05$~T field, is well described by the Curie-Weiss law between 200 and 400~K (shown in the inset), where the filled symbols show the susceptibility and open symbols mark the inverse susceptibility. A sharp cusp at $T_N = 0.78$~K marks an antiferromagnetic ordering transition. \textbf{(c)} The same transition is also detected in heat capacity measurements and the computed entropy release approaches $R\ln{(2)}$ (shown in the inset), as expected for a crystal field ground state doublet.} 
\label{structure}
\end{figure*}

In this work, we investigate the magnetic ground state of a prototype rare-earth triangular lattice antiferromagnet, CuNdO$_2$. This delafossite-structured material possesses an undistorted triangular lattice of Nd$^{3+}$~\cite{miyasaka2009synthesis}. Long-range magnetic order at $T_N=0.78$~K is detected by heat capacity, magnetic susceptibility, and muon spin relaxation, but without accompanying magnetic Bragg peaks in neutron diffraction. Consideration of single-ion anisotropy is the key ingredient to resolve this apparent contradiction: crystal field analysis reveals that Nd has a dominantly Ising character with a very small in-plane component. The extreme frustration of the antiferromagnetically coupled Ising moments is alleviated through the ordering of the significantly smaller in-plane component of the Nd$^{3+}$ moment. Based on linear spin wave modelling of the collective spin excitation, we propose a 120\textdegree\ ordered state for CuNdO$_2$ with a nearly vanishing magnitude of the ordered moment.

\section*{Results and Discussion}

\subsection{Long range antiferromagnetic order}

CuNdO$_2$ crystallizes in the $R\overline{3}m$ delafossite structure, as shown in Fig.~\ref{structure}(a). This structure is composed of staggered layers of edge-sharing NdO$_6$ octahedra, separated by two-dimensional Cu layers, following an $ABC$ stacking sequence. The magnetic Nd$^{3+}$ cations form an undistorted triangular lattice in the $ab$-plane, while Cu$^{+}$ is non-magnetic. The relative in-plane (3.71~\AA) vs. out-of-plane (6.08~\AA) Nd-Nd spacings, as well as the overall expected extent of orbital overlap, make CuNdO$_2$ a good realization of a two-dimensional triangular lattice material.

Magnetic susceptibility measurements on CuNdO$_2$ are shown in Fig.~\ref{structure}(b). Between 200 and 400 K, the susceptibility follows a standard Curie-Weiss law behavior with a fitted paramagnetic moment of 3.69 $\mu_{\text{B}}$ per Nd$^{3+}$, in excellent agreement with the expected Hund’s rules value~\cite{mugiraneza2022tutorial}. Below 200 K, the slope increase in the inverse susceptibility signals the reduction of the Nd$^{3+}$ moment as the excited crystal field levels are thermally depopulated. Below 5~K, the paramagnetic moment is reduced to 1.79 $\mu_{\text{B}}$ per Nd$^{3+}$ and the Curie-Weiss temperature in this fitting range is $-4.0$~K, suggesting dominant antiferromagnetic interactions. At $T_N = 0.78$~K, a sharp cusp is observed, marking an antiferromagnetic ordering transition. Over all measured temperatures, no splitting is observed between measurements performed with and without field cooling.  

Low temperature heat capacity measurements on CuNdO$_2$, shown in Fig.~\ref{structure}(c), confirm the presence of a magnetic ordering transition. A sharp lambda-like anomaly is observed at $T_N = 0.78$~K, in good agreement with previous reports~\cite{miyasaka2009synthesis}. Here, the N\'eel temperature is defined as the maximum in $C/T$. The calculated entropy release associated with this anomaly, shown in the inset of Fig.~\ref{structure}(c), approaches a value of $R\ln{(2)}$ by 4 K, as expected for long-range order arising from a well-isolated ground state doublet.

\begin{figure*}
\centering
\includegraphics[width=\textwidth]{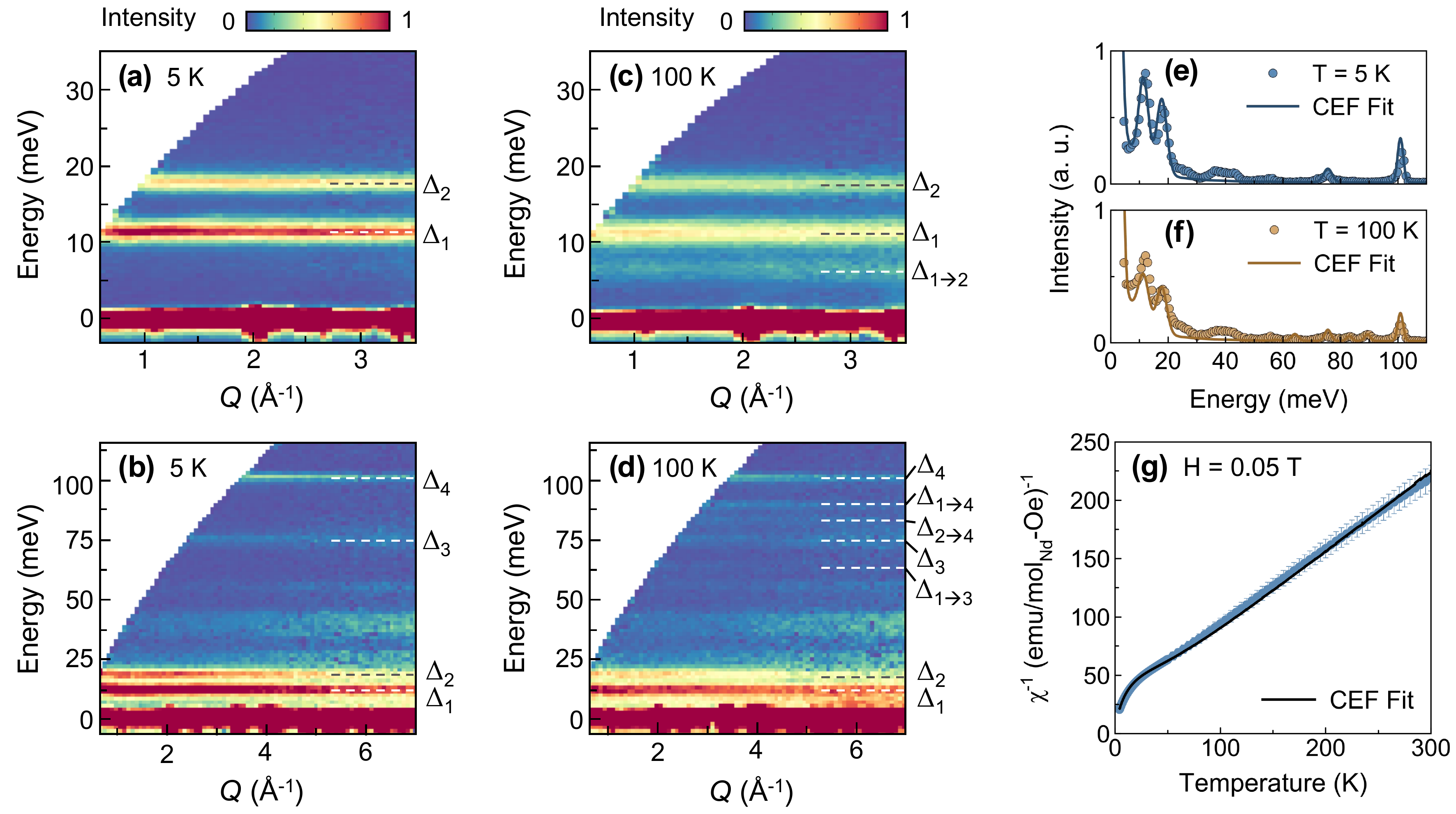}
 \caption{\textbf{The crystal field ground state of Nd$\mathbf{^{3+}}$ in CuNdO$_2$ is an effective spin-$\mathbf{\sfrac{1}{2}}$ with strong Ising anisotropy.} Inelastic neutron scattering measurements performed at \textbf{(a,b)} $T=5$~K and \textbf{(c,d)} $T= 100$~K with $E_i=45$ and 150~meV, respectively. The CEF excitations from the ground state ($\Delta_1$ to $\Delta_4$) and those arising from thermally populated levels ($\Delta_{1\rightarrow2}$, $\Delta_{1\rightarrow3}$, $\Delta_{2\rightarrow4}$, and $\Delta_{1\rightarrow4}$) are indicated by the dashed lines. These experimentally observed CEF levels were used to determine the CEF Hamiltonian, the result of which is shown at \textbf{(e)} $T=5$~K and \textbf{(f)} $T=100$~K. The positions of unmatched intensity at 25, 40, and 55~meV correspond to intense phonon modes. \textbf{(g)} The paramagnetic susceptibility of CuNdO$_2$ computed with the fitted CEF Hamiltonian, showing good agreement with the experimental data.}
\label{fig:CEF}
\end{figure*}

\subsection{Crystal field ground state and single-ion Ising anisotropy}

The single-ion properties of the Nd$^{3+}$ moments in CuNdO$_2$ were characterized using inelastic neutron scattering data. The 5~K and 100~K neutron scattering spectra of CuNdO$_2$ are shown in Fig.~\ref{fig:CEF}(a,b,c,d) for incident neutron beam energies of 45 meV and 150 meV. The spin-orbit ground state manifold of Nd$^{3+}$ with total angular momentum $J=\sfrac{9}{2}$ has a $(2J+1)=10$-fold degeneracy. When subjected to the $D_{3d}$ local crystal electric field (CEF) environment in CuNdO$_2$, this degenerate manifold is expected to split into five Kramers doublets, the lowest energy of which will form the CEF ground state. At low temperatures, where only the ground state is populated, we therefore expect to observe four CEF excitations. 

As indicated in Fig.~\ref{fig:CEF}(a,b), at 5~K we observe four dispersionless magnetic transitions at energy transfers of $\Delta_1=11.3(3)$~meV, $\Delta_2=17.7(3)$~meV, $\Delta_3=74.8(5)~$meV, and $\Delta_4=100.8(5)$~meV each of which is characterized by a decrease in scattered intensity as a function of momentum transfer $\mathbf{Q}$, consistent with the magnetic form factor of Nd$^{3+}$. Upon raising the temperature to 100 K, new CEF \textcolor{black}{transitions appear} at energies of $\Delta_{1\rightarrow2}=\Delta_2-\Delta_1=6.4$~meV, $\Delta_{1\rightarrow3}=\Delta_3-\Delta_1=63.5$~meV, $\Delta_{2\rightarrow4}=\Delta_4-\Delta_2=83.1$~meV, and  $\Delta_{1\rightarrow4}=\Delta_4-\Delta_1=89.5$~meV, as labeled in Fig.~\ref{fig:CEF}(c,d). This new set of excitations is compatible with transitions arising from the thermally occupied first ($\Delta_1$) or second ($\Delta_2$) excited CEF level. The additional excitations centered near 25, 40, and 55 meV are phonon modes as can be deduced based on their increasing intensity as a function of $\mathbf{Q}$. We therefore concluded that the single-ion energy scheme of Nd$^{3+}$ in CuNdO$_2$ consists of four excited states with respective energies of $\Delta_1$, $\Delta_2$, $\Delta_3$, and $\Delta_4$ lying above the CEF ground state.

\begin{figure*}[htbp]
\centering
\includegraphics[width=\textwidth]{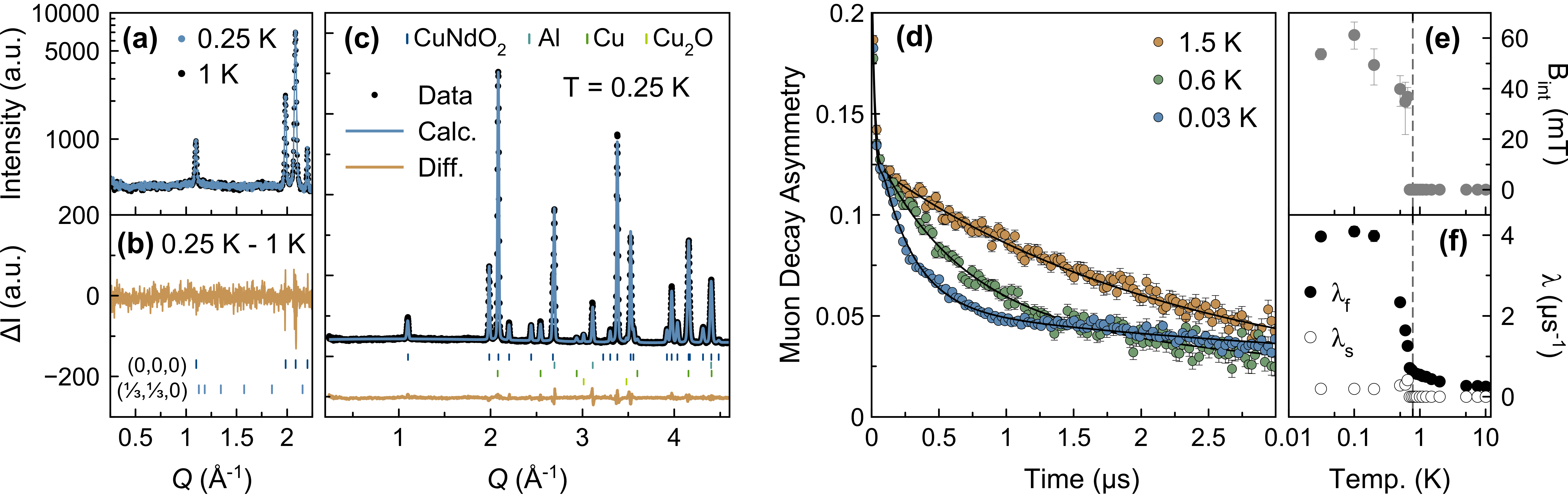}
 \caption{\textbf{Absence of magnetic Bragg peaks in CuNdO$_2$ despite clear bulk ordering observed with $\mu$SR.} \textbf{(a)} Neutron diffraction measurements on CuNdO$_2$, plotted on a log intensity scale, above (1 K) and below (0.25~K) the ordering temperature $T_N = 0.78$~K. \textbf{(b)} Difference of the elastic intensity above and below $T_N$ confirming the absence of any magnetic Bragg peaks within the signal-to-noise ratio of the experiment. \textbf{(c)} Rietveld refinement of the neutron diffraction pattern of CuNdO$_2$ collected at 0.25~K, showing excellent agreement ($\chi^2=1.02$) with a purely structural description. The refined phases include Al, originating from the sample can, and minor Cu and Cu$_2$O impurities, which are residual from the synthesis. \textbf{(d)} Representative muon decay asymmetry spectra at varying temperatures above and below $T_N=0.78$~K. The data are fitted with the function shown in Eq.~(\ref{muon-fcn}). Error bars represent one standard deviation. The temperature dependence of the fitted \textbf{(e)} internal field of the oscillatory component and \textbf{(f)} relaxation rate of the fast (filled symbols) and slow (open symbols) components, consistent with the magnetic transition at $T_N=0.78$~K marked by the dashed line.}
\label{fig:neutron-muon}
\end{figure*}

The experimentally determined CEF energy scheme was then used to determine the CEF Hamiltonian ($\hat{H}_{CEF}$) of Nd$^{3+}$ in CuNdO$_2$. Within the Stevens operator ($\hat{O}^m_n$) formalism, the CEF Hamiltonian of CuNdO$_2$ is given by:
\begin{align}
H_{\text{CEF}} =\ 
&B^0_2 \hat{O}^0_2 + B^0_4 \hat{O}^0_4 + B^3_4 \hat{O}^3_4 \notag \\
&+ B^0_6 \hat{O}^0_6 + B^3_6 \hat{O}^3_6 + B^6_6 \hat{O}^6_6,
\end{align}
which is appropriate for the $D_{3d}$ point-group symmetry of Nd$^{3+}$ in this crystal structure~\cite{Gao2020,Scheie2020}. The Stevens parameters were refined by fitting the observed energies and intensities of all CEF excitations and were further constrained by fitting the paramagnetic susceptibility, as described in the Methods. The best fit to the data was obtained with $B^0_2=0.6934~$meV, $B^0_4=2.92~\mu$eV, $B^3_4~=~-0.30$~meV, $B^0_6=-0.28~\mu$eV, $B^3_6=3.45~\mu$eV, and $B^6_6=-5.50~\mu$eV. The resulting CEF model is compared to the inelastic neutron scattering data in Fig.~\ref{fig:CEF}(e,f) where the 2D energy spectrum at both 5 K and 100 K was obtained by integrating the data in Fig.~\ref{fig:CEF}(b,d) for momentum $Q$ between 3.5 and 5 $\angstrom^{-1}$. Satisfactory agreement between the calculated spectrum and the data is obtained. The unmatched intensity at 25, 40, and 55 meV is due to phonons modes, which are not included in this model.  The fitted CEF Hamiltonian also provides good agreement with the temperature dependence of the magnetic susceptibility (Fig.~\ref{fig:CEF}(g)).  

The resulting CEF ground state of the Nd$^{3+}$ pseudo-spins in CuNdO$_2$ corresponds to:
\begin{align}
| \text{G.S.}\pm \rangle 
= \mp a\, |\pm \tfrac{1}{2} \rangle 
+ b\, |\mp \tfrac{5}{2} \rangle 
+ c\, |\pm \tfrac{7}{2} \rangle
\label{CEF_GS}
\end{align}
with $a=0.434$, $b=0.883$, and $c=0.178$. \textcolor{black}{This Kramers doublet is described by a $\Gamma_4$ irreducible representation so it behaves essentially like a $S=\sfrac{1}{2}$ spin degree of freedom~\cite{Rau2019}} The dipole moment associated with this pseudo-spin-$\sfrac{1}{2}$ CEF ground state is highly anisotropic, with an out-of-plane (Ising) component given by:
\begin{equation} 
\begin{split}
\mu_z = g_J \mu_{\text{B}} |\langle + | J_z | + \rangle| & = \tfrac{1}{2} g_J \mu_{\text{B}} |(-a^2 + 5b^2 - 7c^2)| \\
      & = 1.27(4)\,\mu_{\text{B}}
\end{split}
\end{equation}
and an in-plane (XY) component given by:
\begin{equation} 
\begin{split}
\mu_\perp = g_J \mu_{\text{B}} \left| \langle + | J_+ | - \rangle \right| &= \tfrac{1}{2} g_J \mu_{\text{B}} |(8bc - 5a^2)| \\
          &= 0.11(8)\,\mu_{\text{B}}
\end{split}
\end{equation}
The CEF analysis therefore shows that the Nd$^{3+}$ moments in CuNdO$_2$ have a dominant Ising character (\emph{i.e.} point predominantly out-of-plane). With this understanding of the single ion anisotropy established, we now move on to explore the nature of the magnetic ground state.

\subsection{Magnetic ground state}

\begin{figure*}
\centering
\includegraphics[width=\textwidth]{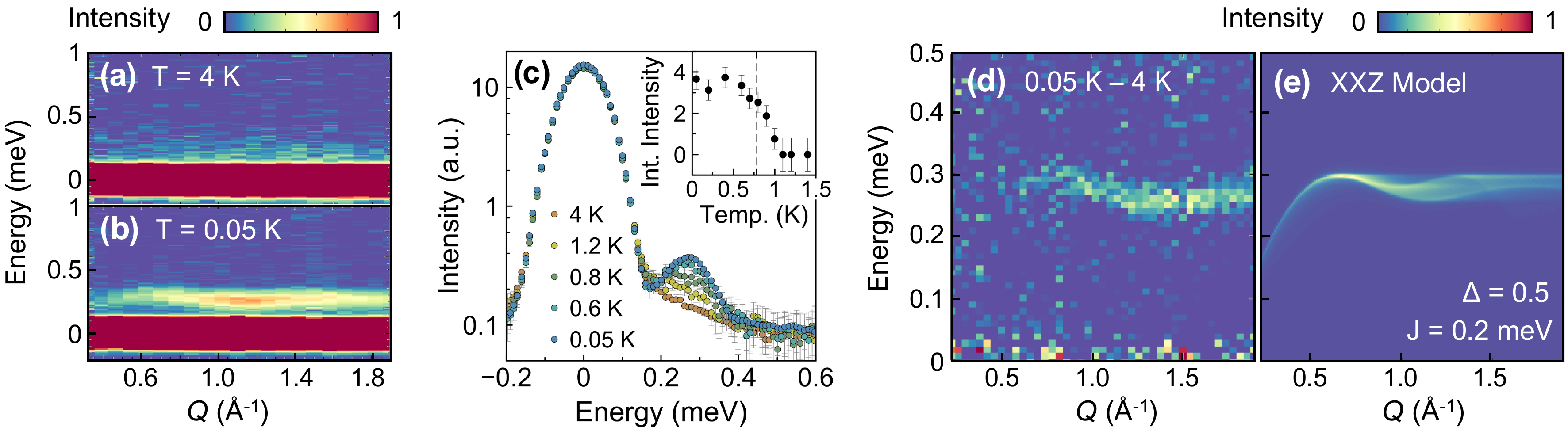}
 \caption{\textbf{Weakly dispersive low-temperature spin excitations in CuNdO$_2$.} Inelastic neutron scattering measurements of CuNdO$_2$ acquired at \textbf{(a)} $T=4$~K and \textbf{(b)} $T=0.05$~K using neutrons with incident energies of $E_i=3.6~$meV ($\lambda = 4.8$~$\angstrom$), revealing the formation of a nearly dispersionless collective spin excitation below $T_N$. \textbf{(c)} Temperature dependence of the $Q$-integrated $4.8~\angstrom$ inelastic neutron scattering spectra. Error bars represent one standard deviation. The inset shows the temperature dependence of the integrated scattered intensity associated with the 0.27(2)~meV inelastic neutron scattering mode. \textbf{(d)} Inelastic neutron scattering spectra of CuNdO$_2$ collected with $E_i=2.27$~meV ($\lambda=6$~$\angstrom$) incident neutrons energy at $0.05$~K. A background measurement acquired at 4~K was subtracted from this data set. \textbf{(e)} Powder-averaged linear spin-wave calculation for CuNdO$_2$ assuming a 120\textdegree\ antiferromagnetic order, which is energetically stabilized by a XXZ spin Hamiltonian with exchange constant $J=0.2$~meV and an anisotropic exchange coefficient $\epsilon=0.5$. }
\label{fig:FOCUS}
\end{figure*}

To better understand the magnetic ground state of CuNdO$_2$, we performed powder neutron diffraction measurements. The neutron wavelength for the measurement, $\lambda = 2.41$~\AA, was selected to maximize our coverage of the smallest momentum transfers, where the magnetic intensity is expected to be highest due to the magnetic form factor. High statistics data sets were collected at 1~K and 0.25~K, which are above and below $T_N=0.78$~K, respectively. Direct comparison of these data sets, which are overplotted in Fig.~\ref{fig:neutron-muon}(a) on a log intensity scale, reveals no changes in Bragg peak intensity nor formation of any new \textcolor{black}{commensurate or incommensurate} magnetic Bragg peaks due to the onset of long-range magnetic order. This absence of magnetic Bragg peaks is reinforced by examining the difference between the two data sets, shown in Fig.~\ref{fig:neutron-muon}(b), showing no deviations from zero within the signal-to-noise of the experiment. 

The neutron diffraction data at both temperatures can be fully captured through a purely structural refinement, as shown in Fig.~\ref{fig:neutron-muon}(c). The complete pattern includes a contribution from the aluminum sample can, as well as two weak non-magnetic impurities, Cu (1\%) and Cu$_2$O (2\%), which are residual from the synthesis. We can therefore conclude that there are no observable magnetic Bragg peaks in the neutron diffraction pattern of CuNdO$_2$ within its ordered state.

Muon spin relaxation measurements on CuNdO$_2$ provide a clear counterpoint to the neutron diffraction result, showing clear evidence of magnetic order. Representative zero field muon decay asymmetry spectra are shown in Fig.~\ref{fig:neutron-muon}(d), where increasing relaxation and structure are evident upon cooling into the ordered state. The data were fitted according to the following function:
\begin{align}
A(t) = A_{samp}[(1-f_F-f_S)e^{-\lambda_Bt}\cos{(\gamma_{\mu}B_{int}t)} \notag \\
+ f_Fe^{-\lambda_Ft}
+f_Se^{-\lambda_St}] + A_{bg},
\label{muon-fcn}
\end{align}
where $A_{bg}$ accounts for the 5\% non-relaxing fraction of asymmetry from muons that land outside the sample, typical for the spectrometer. The remaining 95\% of the asymmetry originates from the sample, $A_{samp}$, and is strongly relaxing, consistent with the entire sample undergoing magnetic order. The sample fraction required three components to achieve a satisfactory fit: one decaying oscillation ($f_{osc} = 1 - f_F - f_S$), a second with a fast relaxation rate ($f_F$), and a third with a slower relaxation rate ($f_S$). 

The temperature dependence of the internal field ($B_{int}$), and the fast and slow relaxation rates ($\lambda_f$ and $\lambda_s$) are presented in Fig.~\ref{fig:neutron-muon}(e,f). All three parameters show a sharp change below $T_N$, in good agreement with the transition established in susceptibility and heat capacity. 
\textcolor{black}{The fitted internal field of the oscillatory component is 60 mT within the magnetically ordered state; however, since no long-lived precession could be observed due to the high damping, there is a large uncertainty associated with the exact value of this parameter. The large damping suggests that the distribution of fields at the muon site is large compared to the average value of approximately 60 mT.}
The fast relaxing component rate follows power law-like behavior below $T_N$, suggesting this component also originates from magnetic order. This component most likely arises from a second, magnetically distinct muon stopping site, and would likely have a second oscillation frequency associated with it. As there is no second oscillation resolvable, we can conclude that the second internal field is small compared to the static distribution of fields at this muon site, which is captured by the large value of $\lambda_F$. This large magnitude of $\lambda_F$ could also indicate the presence of some magnetic disorder.

The component with a slow relaxation rate suggests that some weak dynamics persist below $T_N$, while the dynamics above $T_N$ are captured by the fast relaxation rate. The persistence of the fast relaxation rate above $T_N$ can be understood as originating from muons stopping close to the large Nd$^{3+}$ moments, leading to a large distribution of internal fields (centered around zero). The presence of two components above $T_N$ suggests that CuNdO$_2$ is in the slow fluctuation limit, whereas a normal paramagnet would be in the fast fluctuation regime. This type of behavior is commonly observed for frustrated magnets~\cite{zapf2016magnetization,ziat2017frustrated,mutch2020long}. This analysis is consistent with data collected under weak transverse field conditions, which shows a response typical for a magnetic transition in a powder sample.

In attempting to reconcile the clear signatures of magnetic order observed with magnetometry, heat capacity, and $\mu$SR with the absence of magnetic Bragg peaks observed with neutron diffraction, there are two possible explanations we can consider. The first is that the ordering degrees of freedom are not magnetic dipoles but are instead a higher order multipole~\cite{pourovskii2025hidden}. This ``hidden order'' scenario is precluded by our determination of the Nd$^{3+}$ CEF ground state, which has a definitively dipolar character. Another possibility is that the ordered magnetic moments of the Nd$^{3+}$ ions are below the detection limit of our diffraction experiment. This scenario would be consistent with our CEF analysis, assuming the Nd$^{3+}$ pseudo-spins develop a purely in-plane spin structure below $T_N$. 
In the next section, we explore this scenario by studying the spin dynamics of CuNdO$_2$.


\subsection{Spin wave spectrum}

Our final experimental evidence related to the magnetic ground state of CuNdO$_2$ comes from measurements of its collective spin excitations with inelastic neutron scattering. In contrast to the earlier presented inelastic neutron scattering measurements that were used to establish the CEF scheme, these measurements were performed at lower temperatures and with smaller incident energies. Neutron scattering spectra with an incident neutron beam energy of $3.55$~meV ($\lambda=4.8$~\AA) at a temperature of $4$~K and $0.05$~K, which are respectively above and below $T_N$, are presented in Fig.~\ref{fig:FOCUS}(a,b). Above $T_N$, consistent with the CEF scheme, no inelastic neutron scattering is observed for energy transfers up to 1~meV. Below $T_N$, however, we observe a weakly dispersive excitation centered at an energy transfer close to 0.3~meV, which signifies a symmetry breaking of the CEF ground-state doublet. The intensity of this excitation is observed to peak close to 1.1~\AA$^{-1}$.

To investigate the thermal evolution of this spin excitation, we collected identical spectra for temperatures between $0.05$~K and $4$~K. The $Q$-integrated spectra (Fig.~\ref{fig:FOCUS}(c)) show that an inelastic peak is clearly resolved only below approximately $1$~K, while it merges with the quasi-elastic scattering at higher temperatures. To quantify the temperature dependence, we subtracted the $4~$K data set from each lower temperature data set and fitted the resulting inelastic peak to a Gaussian profile of variable area. The temperature dependence of this integrated area, shown in the top right inset of Fig.~\ref{fig:FOCUS}(c), indicates that the inelastic peak develops within the critical thermal region of the magnetic ordering at $T_N = 0.78$~K.   

Given the weakly dispersive nature of the observed spin excitation in CuNdO$_2$, we conducted inelastic neutron scattering measurements with enhanced energy resolution using $E_i=2.27$~meV ($\lambda=6$~\AA) incident neutrons. The resulting 0.05~K spectrum, after subtracting a 4~K background, is shown in Fig.~\ref{fig:FOCUS}(d). A Gaussian fit to the data, integrated over momentum transfers $Q=[1,1.2]~\angstrom^{-1}$, shows the magnon mode is centered at $0.27(2)$~meV with a full width at half maximum (FWHM) of 0.06(1)~meV, which is comparable to the instrumental resolution of 0.04~meV. While neutron absorption effects strongly suppress the low $Q$ scattering, we can resolve weak dispersion in this higher resolution measurement. 


Despite the absence of resolvable magnetic Bragg peaks, this measurement of the collective spin dynamics allows us to identify the likely magnetic ground state of CuNdO$_2$. We consider a minimal model of nearest-neighbor XXZ spins sitting on a triangular lattice: 
\begin{equation}
    H=J \sum_{\langle i j\rangle}\left(S_{i}^{x} S_{j}^{x}+S_{i}^{y} S_{j}^{y}+\epsilon S_{i}^{z} S_{j}^{z}\right).
    \label{eq:XXZ}
\end{equation}
\textcolor{black}{Here $S$ denotes the pseudo-spin operators associated with the ground-state Kramers doublet in Eq.~(\ref{CEF_GS}),} and $J$ represents the overall strength of the exchange interaction, while $\epsilon$ characterizes the anisotropy between in-plane and out-of-plane pseudo-spin components. \textcolor{black}{In this model, we assume the system is projected onto its CEF ground-state doublet, an approximation well justified by the fact that the ordering temperature of CuNdO$_2$ ($T_N=0.78$~K) is nearly three orders of magnitude smaller than the energy gap to the first excited CEF level ($\sfrac{\Delta_1}{k_B} = 131$~K).} A significantly anisotropic value of $\epsilon$ is expected, as the $g$-factor anisotropy originating from the crystal field ground state wavefunction naturally leads to anisotropic spin exchange interactions as well.

To fit the inelastic neutron scattering data, we employ linear spin wave theory parameterized by $J$ and $\epsilon$ (see Methods for details). We find that the best fit to the data is obtained with $J = 0.2$~meV and $\epsilon = 0.5$, the result of which is shown in Fig.~\ref{fig:FOCUS}(e). These parameters indicate that the magnetic ground state of CuNdO$_2$ is a 120\textdegree\ antiferromagnetic order on the triangular lattice, with the ordered moments confined to the $ab$-plane. 
The small in-plane $g$-factor, previously determined in our CEF analysis, significantly suppresses the dipole matrix elements for elastic neutron scattering, leading to an undetectably small magnetic Bragg intensity despite the presence of long-range magnetic order. The wavevector positions for this ordered state, which has a propagation vector of $k=(\sfrac{1}{3},\sfrac{1}{3},0)$, are noted in Fig.~\ref{fig:neutron-muon}(b). \textcolor{black}{Our neutron diffraction refinement indicates that a 120\textdegree\ order with Nd$^{3+}$ ordered moments smaller than 0.3~$\mu_{\text{B}}$ is unresolvable, which is a value greater than the in-plane CEF ground-state moment, $\mu_\perp = 0.11(8)~\mu_{\text{B}}$.} This ordered state then accounts for the absence of a measurable dipole moment in neutron diffraction experiments. 

The nearly dispersionless nature of the magnon modes observed in inelastic neutron scattering can be attributed to the reduced Ising (out-of-plane) exchange anisotropy, $\epsilon < 1$, which limits the bandwidth of the magnon branch, effectively flattening it in energy. Thus, the minimal XXZ model with anisotropic exchange together with the small in-plane $g$-factor successfully captures the key experimental observations, including the magnetic long-range order found by specific heat and $\mu$SR, the absence of a detectable dipole ordered moment with neutrons, and the presence of seemingly dispersionless magnon excitations.  


In summary, we synthesized and thoroughly characterized the collective magnetism of the rare-earth delafossite CuNdO$_2$, which hosts a frustrated triangular lattice of highly anisotropic pseudo-spins $S=\sfrac{1}{2}$. We found these pseudo-spins have a strong out-of-plane Ising anisotropy with a vanishing in-plane dipole moment of 0.11(8)~$\mu_{\text{B}}$. Despite this strong Ising anisotropy, the magnetic frustration inherent to the triangular lattice is relieved via an in-plane ordering into a 120\textdegree\ spin structure. While the resulting neutron Bragg diffraction of this antiferromagnetic structure is too weak to be resolved in our diffraction experiment, clear signatures of long-range magnetic order are found in the magnetic susceptibility, heat capacity, $\mu$SR, and inelastic neutron scattering measurements. We expect this phenomenology to commonly appear in rare-earth frustrated magnets, especially ones with strong single-ion anisotropy, where the easy and hard axes respectively promote and relieve magnetic frustration.  



\section*{Methods}

\subsection*{Synthesis} 
Polycrystalline samples of CuNdO$_2$ were synthesized by solid state reaction of Cu$_2$O and Nd$_2$O$_3$. A small excess (5\% by stoichiometry) of Cu$_2$O was needed to achieve a complete reaction of all Nd$_2$O$_3$. The reagents were ground in a mortar and pestle, pressed into a pellet, and reacted at 900\textdegree C for 16 h. Three additional heating cycles of 72 h at 875\textdegree C were required to achieve a complete reaction, with intermediate re-grinding and re-pelleting each time. As Cu-based delafossites are known to be tolerant of high levels of oxygen doping~\cite{cava1993lacuo25+,isawa1997synthesis}, the stoichiometric sample was prepared under mildly reducing conditions by using a rotary pump to achieve active vacuum throughout the synthesis. Powder x-ray diffraction was performed to confirm the crystalline phase of the resulting sample with a Bruker D8 Advance. The final sample is free from any magnetic impurities but has a small amount of residual non-magnetic Cu$_2$O (2\%) and Cu (1\%).

\subsection*{Bulk characterization} 
Magnetometry measurements were performed on a Quantum Design Magnetic Property Measurement System (MPMS) equipped with a $^3$He insert, giving a base temperature of 0.4~K. Susceptibility was measured with an applied field of $\mu_0H = 0.05$~T. Heat capacity measurements were performed on a Quantum Design Physical Property Measurement System (PPMS) equipped with a dilution refrigerator insert, giving a base temperature of 0.2~K. 

\subsection*{Neutron diffraction} 
Powder neutron diffraction was performed on the HB-2A beamline at the High Flux Isotope Reactor (HFIR) at Oak Ridge National Lab (ORNL). A 2-gram sample of CuNdO$_2$ was loaded in an aluminum sample can with an over-pressure of He (10~atm). A base temperature of 0.25~K was achieved using a dry Oxford Instruments’ Heliox-7 system. High statistics (12 hour) scans were collected with a neutron wavelength of $\lambda = 2.41$~\AA\ above (1~K) and below (0.25~K) the magnetic ordering transition. The diffraction patterns were refined with FullProf~\cite{rodriguez1993recent} and representation analysis for candidate magnetic structures was performed with SARAh~\cite{wills2000new}.  

\subsection*{Muon spin relaxation} 
Muon spin relaxation ($\mu$SR) measurements were performed on the FLAME spectrometer at the Swiss Muon Source, Paul Scherrer Institute. A sintered pellet of CuNdO$_2$ crushed into multiple pieces was mounted sandwiched between two 25 micron thick copper foils filling an area of approximately 1~cm$^2$. The sample was cooled to a base temperature of 30~mK using a dilution refrigerator. Measurements were collected under zero field, weak transverse field, and longitudinal field conditions. 

\subsection*{Inelastic neutron scattering} 
The crystal electric field (CEF) levels of CuNdO$_2$ were measured on the SEQUOIA time-of-flight spectrometer at ORNL~\cite{granroth2010sequoia}. Measurements were performed on a 2-gram sample at 5~K and 100~K with incident neutron energies of 45~meV and 150~meV. The high-resolution Fermi chopper was used for both configurations with a frequency of 360~Hz and 600~Hz for the 45~meV and the 150~meV configurations, respectively. Furthermore, we used the same T$_0$ chopper frequency of 90~Hz for both configurations.

The CEF Hamiltonian was parameterized within the Stevens operator formalism based on the experimentally observed CEF energies and intensities and further constrained by the paramagnetic susceptibility. The Stevens parameters $B^m_n$ were initially estimated using the PyCrystalField software~\cite{Scheie2021} and then further refined following a protocol similar to Ref.~\cite{Gaudet2018,Plessis2019}. The temperature dependence of the powder-averaged magnetic susceptibility  was calculated using the single-ion susceptibility~\cite{jensen1991}:
\begin{align}
\chi_{\text{ion}} &= \frac{N_A g_J^2 \mu_{\text{B}}^2}{k_B \sum_n e^{-E_n/T}} \Bigg(
\frac{\sum_n \langle n | \mathbf{J} | n \rangle^2 e^{-E_n/T}}{T} \notag \\
&\quad + \sum_{n} \sum_{\substack{m \\ m \neq n}} 
\frac{|\langle n | \mathbf{J} | m \rangle|^2}{E_m - E_n} 
(e^{-E_n/T} - e^{-E_m/T})
\Bigg)
\end{align}
where $N_A$ is Avogadro's number, $g_J=0.7273$ is the Land\'e $g$-factor for Nd$^{3+}$, $\mu_{\text{B}}$ is the Bohr magneton, $k_B$ is the Boltzman constant, $\mathbf{J}$ is the total angular momentum operator, and $E_n$ is the calculated energy of the $n$th CEF level. The diamagnetic susceptibility ($\chi_{dia}$) and correlation effects were estimated by fitting the susceptibility data to:
\begin{align}
\chi=\chi_{dia} + \frac{\chi_{ion}}{1+\lambda \chi_{ion}}
\end{align}
where $\lambda$ is the Weiss molecular field constant. 

The low energy spin excitation spectrum of CuNdO$_2$ was measured at the FOCUS time-of-flight spectrometer at the Swiss Spallation Neutron Source (SINQ) at the Paul Scherrer Institute (PSI)~\cite{mesota1996focus}. We collected data using both $E_i=3.55~$meV and $E_i=2.27~$meV, yielding energy resolutions at the elastic line of 0.08 and 0.04~meV, respectively. The 2-gram powder sample was inserted into a Cu sample can filled with a 10 bar He-gas atmosphere to ensure good thermal contact. The sample can was inserted into a dilution fridge with a minimum temperature of 50~mK.

\subsection*{Modelling} 
To understand the magnon dispersion, we performed linear spin wave theory calculations on the XXZ model:
\begin{equation}
    H=J \sum_{\langle i j\rangle}\left(S_{i}^{x} S_{j}^{x}+S_{i}^{y} S_{j}^{y}+\epsilon S_{i}^{z} S_{j}^{z}\right).
    \label{eq:XXZ}
\end{equation}
This model is known to host a $\mathrm{120^\circ}$ ground state when $\epsilon \in [-0.5, 1.0]$ ~\cite{momoiGroundStatePropertiesPhase1992}. 
To find the magnon band dispersion, we apply a standard Holstein-Primakoff transformation:
\begin{equation}
    \begin{aligned}
        S_{i,d}^{+'} &= \sqrt{2S - d_i^\dagger d_i} d_i\\
        S_{i,d}^{-'} &= d_i^\dagger \sqrt{2S - d_i^\dagger d_i} \\
        S_{i,d}^{z'} &= S - d_i^\dagger d_i,
    \end{aligned}
\end{equation}
where $d=a,b,c$ refers to the three sub-lattices, and $z'$ is aligned to the local spin directions. \textcolor{black}{Here $S^\pm$  raises/lowers the total angular momentum by $\pm 1$. Excitations to higher crystal-field levels, such as the one at $\Delta_1 = 11$ meV containing $|\pm \tfrac{3}{2} \rangle$, are neglected in this analysis, since their contributions are of a higher order, which lies beyond the linear spin wave theory.} In the $\mathrm{120^\circ}$ order, these coordinates are related by $C_3^{(c)}$ rotation. The Hamiltonian would transform accordingly:
\begin{equation}
    \begin{aligned}
        H &= - \frac{J}{2} \sum_{\braket{ij}} \left( S_i^{z'} S_j^{z'} + S_i^{x'} S_j^{x'} - 2\epsilon S_i^{y'} S_j^{y'} \right) \\
        &\phantom{=}+ \frac{\sqrt{3}J}{2} \sum_{(i \to j)} \left( S_i^{z'} S_j^{x'} - S_i^{x'} S_j^{z'} \right)
    \end{aligned}
\end{equation}.

Then by expanding $H$ to the quadratic order of Holstein-Primakoff bosons, one obtains:
\begin{equation}
    H = E_0 + \sum_{\mathbf{k}}\mathcal{D}_{\mathbf{k}}^\dagger \mathcal{H}_{\mathbf{k}} \mathcal{D}_{\mathbf{k}} 
\end{equation}
where $E_0 = -(3/2)JN_{\text{site}} S(S+1)$ is the ground state's energy, $\mathcal{D}_\mathbf{k}^\dagger = (\mathcal{A}_{\mathbf{k}}^\dagger, \mathcal{A}_{-\mathbf{k}}^T)$ and $\mathcal{A}_{\mathbf{k}}^\dagger = (a_{\mathbf{k}}^\dagger, b_{\mathbf{k}}^\dagger, c_{\mathbf{k}}^\dagger )$ contains all the bosons.

$\mathcal{H}_{\mathbf{k}}$ can be written as:
\begin{equation}
    \mathcal{H}_{\mathbf{k}} = \frac{3JS}{2}\left[ \mathbb{I}_{2\times 2} \otimes(\mathbb{I}_{3\times 3} - \mathcal{M}_{\mathbf{k}}^- ) - \sigma_x \otimes \mathcal{M}_{\mathbf{k}}^+\right],
\end{equation}
where:
\begin{equation}
    \mathcal{M}_{\mathbf{k}}^\pm = \frac{1}{4}\begin{pmatrix}
        0 & (1 \pm 2\epsilon) \gamma_{\mathbf{k}} & (1 \pm 2\epsilon) \gamma_{\mathbf{k}}^*\\
        (1 \pm 2\epsilon) \gamma_{\mathbf{k}}^* & 0 & (1 \pm 2\epsilon) \gamma_{\mathbf{k}}\\
        (1 \pm 2\epsilon) \gamma_{\mathbf{k}} & (1 \pm 2\epsilon) \gamma_{\mathbf{k}}^* & 0
    \end{pmatrix}.
\end{equation}
Here $\gamma_k = (1/3) \sum_{j=1}^{3} \exp{(\mathrm{i} \mathbf{k}\cdot\delta_j)}$, where $\delta_1 = a_0(1,0)$, $\delta_2 = a_0(-\frac{1}{2}, \frac{\sqrt{3}}{2})$, and $\delta_3 = a_0(-\frac{1}{2}, -\frac{\sqrt{3}}{2})$ are three bond vectors. Its real part $\mathrm{Re}[\gamma_{\mathbf{k}}] = (1/6) \sum_\delta \exp{(\mathrm{i} \mathbf{k}\cdot\delta)} $ is a more common form of $\gamma_{\mathbf{k}}$ when the unit cell is not enlarged. Diagonalizing the above Hamiltonian yields:
\begin{equation}
    \omega_{n,\mathbf{k}} = \frac{3JS}{2}\sqrt{ \Big(1-\mathrm{Re}[\mathrm{e}^{\mathrm{i} \frac{2n\pi}{3}}\gamma_{\mathbf{k}}] \Big) \Big(1 + 2\epsilon\mathrm{Re}[\mathrm{e}^{\mathrm{i} \frac{2n\pi}{3}}\gamma_{\mathbf{k}}]\Big) }.
\end{equation}

In the original Brillouin zone these three bands are related by translations: $\delta\mathbf{k}=a_0^{-1}(\pm4\pi/3,0)$, which corresponds to a shift from $\Gamma$ to $K$ and $K'$, and at each of these points we have a gapless node. In the reduced Brillouin zone, however, $\delta\mathbf{k}$ now becomes the shift from $\Gamma$ to $\Gamma'$, and we only have one gapless node at $\Gamma$ point, which corresponds to the Goldstone mode associated with $U(1)$ symmetry left in the XXZ model. Numerical calculations were performed using the Sunny package~\cite{dahlbom2025sunny}. 
The results show that the magnon bands with $J = 0.2$~meV and $\epsilon \approx 0.5$ best reproduces the key experimental features.\\


\section*{Data availability}
The data that support the findings of this study are provided in the main text. The raw data is available from the corresponding author upon request.

\section*{Acknowledgments}
This work was supported by the Natural Sciences and Engineering Research Council of Canada (NSERC), the Canadian Institute for Advanced Research (CIFAR), and the Sloan Research Fellowships program. This research was undertaken thanks in part to funding from the Canada First Research Excellence Fund, Quantum Materials and Future Technologies Program. HYK acknowledges support from NSERC Discovery Grant No. 2022-04601 and the Canada Research Chairs Program No. CRC-2019-00147.
The support for neutron scattering was provided by the Center for High-Resolution Neutron Scattering, a partnership between the National Institute of Standards and Technology and the National Science Foundation under Agreement No. DMR-2010792. A portion of this research used resources at the High Flux Isotope Reactor and the Spallation Neutron Source, which are DOE Office of Science User Facilities operated by the Oak Ridge National Laboratory. The beam time was allocated to the SEQUOIA and HB-2A instruments on proposal numbers IPTS-27501 and IPTS-27487, respectively. The identification of any commercial product or trade name does not imply endorsement or recommendation by the National Institute of Standards and Technology. This work is partially based on experiments performed at the Swiss Muon Source SµS, Paul Scherrer Institute, Villigen, Switzerland.  \\

\section*{Author contributions}
The sample was synthesized and characterized by D.R. and A.M.H. The neutron diffraction experiment was performed by D.R., A.A.A., S.A.C., and analyzed by A.M.H. Muon spin relaxation experiments were performed by D.R., T.J.H., J.A.K., and H.L., with analysis by T.J.H. Inelastic neutron scattering experiments were performed by J.G., D.R., J.P.E., M.B.S., and A.M.H. Analysis of the crystal electric field ground state was performed by J.G., B.W., and H.Y.K. Spin wave modelling was performed by B.W. and H.Y.K. The manuscript was written by J.G., B.W., H.Y.K., and A.M.H. All authors discussed the results and commented on the manuscript.

\section*{Competing interests}
Hae-Young Kee is an Associate Editor of npj Quantum Materials. Hae-Young Kee was not involved in the journal’s review of, or decisions related to, this manuscript. All other authors declare no competing interests.

\bibliography{bibliography}

\end{document}